\documentclass[showpacs,preprintnumbers,amsmath,amssymb,superscriptaddress]{revtex4}

\usepackage{graphicx}
\usepackage{dcolumn}
\usepackage{bm}
\newcommand{\bra}[1]{\left(#1\right)}

\begin{document}

\preprint{Physica A \textbf{387}, (2008) 4411 - 4416}

\title{Topology and Dynamics of Attractor Neural Networks: The Role of Loopiness}
\author{Pan Zhang}
\affiliation{Institute of Theoretical Physics, Lanzhou University, Lanzhou 730000, China}

\author{Yong Chen}
\altaffiliation{Author to whom correspondence should be addressed. Email: ychen@lzu.edu.cn}
\affiliation{Institute of Theoretical Physics, Lanzhou University, Lanzhou 730000, China}

\date{\today}

\begin{abstract}
We derive an exact representation of the topological effect on the
dynamics of sequence processing neural networks within
signal-to-noise analysis. A new network structure parameter,
loopiness coefficient, is introduced to quantitatively study the
loop effect on network dynamics. The large loopiness coefficient
means the large probability of finding loops in the networks. We
develop the recursive equations for the overlap parameters of neural
networks in the term of the loopiness. It was found that the large
loopiness increases the correlations among the network states at
different times, and eventually it reduces the performance of neural
networks. The theory is applied to several network topological
structures, including fully-connected, densely-connected random,
densely-connected regular, and densely-connected small-world, where
encouraging results are obtained.

\end{abstract}


\maketitle

Starting from pioneering works of modelling complex networks \cite{Watts98,Ba99}, research area related with complex networks have been growing very fast. In parallel to the studies of structural properties of complex networks, there has also been a growing interest in dynamic systems defined on networks. For example, synchronization and collective dynamics, epidemic spreading, cascading failures, opinion formation as well as various strategic games, and some physical model like Ising model (see \cite{boccalettia06} and refs. therein).

Neural assemblies (i.e. local networks of neurons transiently linked by selective interactions) are considered to be largely distributed and linked to form a web-like structure of the brain. Many researches suggested that neural connectivity is far more complex than the random graph. The cortical neural networks of Caenorhabditis elegans and cat were reported to be small-world (SW) and scale-free (SF), respectively \cite{Watts98,nncomplex1}. It is very important to understand how the complex neural wiring architecture is related to brain functions. With same average connections, Hopfield network with random topology was reported to be more efficient for storage and recognition of patterns than both SW network and regular network \cite{Hopfield1982,Patrick03}. For the SF connection, with the same number of synapses, Torres et al. found that the capacity is larger than the storage of the highly diluted random Hopfield networks \cite{Torres03}. Using a Mente-Carlo method to lower the clustering coefficient smoothly with the degree of each neuron kept unchanged, Kim found that the networks with the lower clustering exhibit much better performance \cite{Kim04}.

It is well known that the equilibrium properties of fully-connected Hopfield networks had been extensively studied using spin-glass theory, especially the replica method \cite{replicas,Amit85a,Amit85b}. And the dynamics of fully-connected Hopfield model with static patterns and sequence patterns were widely studied using generating functional analysis \cite{Coolen2000,Coolen98}. As a simple relaxation of biological unrealistical fully-connected model, various random diluted model were studied, include extreme diluted model \cite{Derrida87,Watkin91}, finite diluted model \cite{zhang07,Theumann03}, and finite connection model \cite{WC03}. However, in the case of complex network topology, as far as we know, there are hardly any theoretical studies for dynamics or statics.

In this paper, we use signal-to-noise analysis to study the effect of topology on transient dynamics of sequence processing neural networks. Considering the mathematical convenience, here we only focus on the sequence processing models and the relationship between our results and Hopfield models will be discussed.

The topological effect on neural networks mainly come from loops of
topology \cite{Barkai90}. In the case of so-called extremely diluted
structure ($\lim_{N\to \infty}\bar k^{-1}=\lim_{N\to \infty}\bar
k/N=0$), average loop length is very big, like $\log_{k-1}N$, so
number of short loops (like triangles, quadrangles) in networks will
be very small and effect caused by short loops can be neglected.
Dynamics of networks in this case is easy to study because each spin
at different time steps is uncorrelated \cite{Derrida87}. In
contrast, if such small loops exist, the correlations and feedbacks
in network will lead to more complicated dynamics.

\begin{figure}[h]
\includegraphics[width=0.7\textwidth]{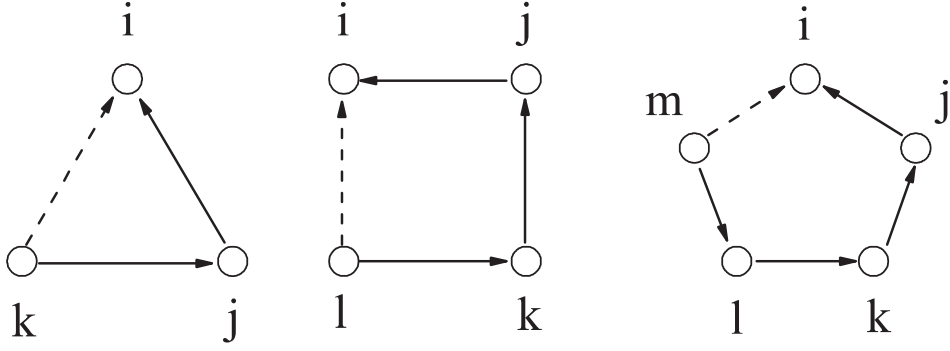}
\caption{Left: the first-order loopiness coefficient, $L_1$, means the probability of connectivity between $k$ and $i$ when $j$ is connected to $i$ and $k$ is connected to $j$. Middle: the second-order loopiness coefficient $L_2$. Right: the third-order loopiness coefficient $L_3$.} \label{fig1}
\end{figure}

In order to quantitatively present the effect from loops, we define a new parameter, loopiness coefficient, to present the probability of finding loops in the network. The definition is shown in Fig. \ref{fig1}. In the left triangle, spin $j$ is connected to spin $i$, and spin $k$ is connected to $j$. We define the probability that $k$ is connected to $i$ as the first-order loopiness coefficient, $L_1$. Similarly, the $n$th-order loopiness coefficient, $L_n$, denotes the linking probability between two vertices to form a loop with $n+2$ edges.

We now study a sequence processing model consisting of $N \rightarrow \infty$ Ising-type neurons $s_i\bra{t} \in \{ +1,-1 \}$. The neurons update their states simultaneously, with the following probabilities,
\begin{equation}
\mathrm{Prob} [s_i \left(t+1\right) | h_i\left(t\right) ] = \frac{e^{\beta s_i\bra{t+1}h_i\bra{t}}}{2\cosh\bra{\beta h_i\bra{t}}} ,
\label{eq01}
\end{equation}
where the local field $h_i\bra{t} = \sum_{j=1}^N J_{ij} s_j\bra{t}$, and $\beta$ is the inverse temperature. For the transfer function $g \bra{\cdot}$, we denote by $s_i \bra{t+1} = g(h_i\bra{t})$.

Let us store $p = \alpha N$ random patterns $\xi^\mu =
\bra{\xi_1^\mu,\ldots,\xi_N^\mu}$ in network, where $\alpha$ is the
loading ratio. So, the interaction matrix $J_{ij} =
\frac{c_{ij}}{Nc} \sum_{\mu=1}^p \xi_i^{\mu+1} \xi_j^\mu$ is chosen
to retrieve the patterns as $\xi^1 \rightarrow \xi^2 \rightarrow
\cdots \rightarrow \xi^p \rightarrow \xi^1$ (note that $\xi^{p+1} =
\xi^1$), where $c_{ij}$ is the adjacency matrix ($c_{ij}=1$ if $j$
is connected to $i$, $c_{ij}=0$ otherwise) \cite{Sompolinsky86}.
Consequently, the degree of spin $i$ is $k_i = \sum_{j \in T_i}
c_{ij} = \sum_{j} c_{ij}$, where $T_i$ is the set of spin $j$
connected with $i$. In this work, for studying the role of
loopiness, we only consider $k_i \approx \bar{k} = Nc$. We assume
that this property holds for dense connected random network, dense
connected
 SW network, and dense connected  regular network.

For any pattern $\xi ^{\nu}$, the order parameter is $m^{\nu} \bra{t} = \frac{1}{N} \sum  \xi _{i}^{\nu} s_{i}\bra{t}$ which represent the overlap between $\mathbf{s} \bra{t}$ and the condensed pattern $\xi ^{\nu}$. The local field in neuron $i$ is described by
\begin{equation}
h_{i}\bra{t} =\frac{1}{Nc} \sum_{j \in T_i} \xi_{i}^{\nu+1} \xi_j^{\nu} s_j\left(t\right) + Z_{i} \left(t\right),
\label{eq02}
\end{equation}
where the first term of the left-hand side is the signal term and the second one the crosstalk noise from uncondensed patterns,
\begin{equation}
Z_i\left(t\right) = \frac{1}{Nc} \sum_{\mu \neq \nu} \sum_{j\in T_i} \xi_i^{\mu+1} \xi_j^\mu s_j\left(t\right).
\label{eq03}
\end{equation}
The state of the spins $s_i(t)$ is determined by the sign of local field $h_i(t)$, that is,
\begin{eqnarray}
s_j \left( t \right) &=& g\left(h_j\bra{t-1}\right) = g \left( \frac{1}{Nc} \sum_{k\in T_j} \xi_j^{\nu} \xi_k^{\nu-1} s_k \bra{t-1} \right. \nonumber\\
&+& \left . \frac{1}{Nc} \sum_{k\in T_j} \xi_j^{\mu} \xi_k^{\mu-1}
s_k\bra{t-1} + \frac{1}{Nc} \sum_{\chi\neq \nu , \mu} \sum_{k\in
T_j} \xi_j^{\chi} \xi_k^{\chi-1} s_k\bra{t-1} \right ). \label{eq04}
\end{eqnarray}
Expanding $s_j\left(t\right)$ into first order, we obtain
\begin{eqnarray}
s_j\left(t\right) &=& \hat{s}_j^\mu\left(t\right) + \frac{1}{Nc} \left . \frac{\partial g}{\partial h} \right |_{h_j\left(t-1\right)} \sum_{k\in T_j} \xi_j^\mu\xi_k^{\mu-1} s_k\left(t-1\right) \nonumber\\
&+& \mathcal{O}\left(N^{-1}\right),
\label{eq05}
\end{eqnarray}
where
\begin{eqnarray}
\hat{s}_j^\mu\left(t\right)&=& g \left( \frac{1}{Nc} \sum_{k\in T_j}
\xi_j^{\nu} \xi_k^{\nu-1} s_k \bra{t-1}+ \frac{1}{Nc} \sum_{\chi\neq
\nu , \mu} \sum_{k\in T_j} \xi_j^{\chi} \xi_k^{\chi-1} s_k\bra{t-1}
\right )\nonumber\\ &=& g\left(h_j\bra{t-1} - \frac{1}{Nc}
\sum_{k\in T_j} \xi_j^{\mu} \xi_k^{\mu-1}
s_k\left(t-1\right)\right).
\end{eqnarray}
Note that, in here, $\hat{s}_j^\mu\left(t\right)$ is independent of
$\xi_j^\mu$. Then the crosstalk noise can be presented by
\begin{eqnarray}
&&Z_{i}\left( t\right) = \sum_{\mu \neq \nu} \frac{1}{Nc} \sum_{j\in T_{i}} \xi _{i}^{\mu +1} \xi _{j}^{\mu } \hat{s}_{j}^{\mu }\left( t\right) \nonumber \\
&+& U\left( t-1\right) \frac{1}{Nc} \sum_{j\in T_{i}} \frac{1}{Nc} \sum_{\mu \neq \nu} \sum_{k\in T_{j}} \xi_{i}^{\mu +1} \xi_{k}^{\mu -1} s_{k}\left(t-1\right),
\label{eq06}
\end{eqnarray}
where $U(t-1) = \langle\!\langle \langle \frac{\partial g} {\partial h} \vert _{h(t-1)} \rangle_{\boldsymbol{h}} \rangle \!\rangle$. $\langle .\rangle _{{\boldsymbol{h}}}$ denotes the average over the distribution of the local field, and  $\left\langle \!\left\langle \cdot \right\rangle \!\right\rangle $ means the average over the initial conditions and the condensed pattern.

Using the central limit theorem, the first term of Eq. (\ref{eq06}) converge to a zero mean Gaussian form $\mathcal{N} (0,\alpha / c )$. Since $\xi_{i}^{\mu +1} \xi _{k}^{\mu-1} s_{k}\left( t-1\right)$ is not independent, we cannot apply this theorem directly into the second term of Eq. (\ref{eq05}), we have to expand $Z_i\bra{t}$ into time $t-2$,
\begin{widetext}
\begin{eqnarray}
Z_{i}\left( t\right) &=&\mathcal{N}(0,\alpha / c ) + U\left( t-1\right) \frac{1}{Nc} \sum_{j\in T_{i}} \frac{1}{Nc} \sum_{\mu \neq \nu} \sum_{k\in T_{j}} \xi _{i}^{\mu +1} \xi _{k}^{\mu -1} \hat{s}^{\mu -1}_{k}\left( t-1\right) \nonumber\\
&&+ U\left( t-1\right) U\bra{t-2} \frac{1}{Nc} \sum_{j\in T_{i}} \frac{1}{Nc} \sum_{k\in T_{j}} \frac{1}{Nc} \sum_{\mu \neq \nu} \sum_{l\in T_{k}} \xi _{i}^{\mu +1} \xi _{l}^{\mu -2} s_{l} \left( t-2\right),
\label{eq07}
\end{eqnarray}
\end{widetext}
with $U(t-2) = \langle\!\langle \langle \frac{\partial g} {\partial h}\vert _{h(t-2)}\rangle_{\boldsymbol{h}} \rangle\! \rangle$. Even though averaging over $j\in T_{i}$, not all $k\in T_{j}$ are residual spins. Note that only $k$ belonging to both $T_{i}$ and $T_{j}$ will survive the averaging operations. The probability of k is described by the first order loopiness coefficient $L_1$ (See Fig. \ref {fig1}). Similarly, from the central limit theorem, it was found that the second term of Eq. (\ref{eq06}) converged to $\mathcal{N}(0,L_{1}U^{2}\left( t-1\right) \alpha / c )$. Then we expand $Z_i\bra{t}$ to time $t-3$,
\begin{widetext}
\begin{eqnarray}
Z_{i}\left( t\right) &=& \mathcal{N} (0,\alpha / c ) + \mathcal{N}(0,L_1 U^2\bra{t-1} \alpha / c ) \nonumber\\
&& + U\left( t-1\right) U\bra{t-2} \frac{1}{Nc} \sum_{j\in T_{i}} \frac{1}{Nc} \sum_{k\in T_{j}} \frac{1}{Nc} \sum_{\mu \neq \nu} \sum_{l\in T_{k}} \xi_{i}^{\mu +1} \xi_{l}^{\mu -2} \hat{s}^{\mu-2}_{l}\left( t-2\right) \nonumber\\
&& + U\left( t-1\right) U\bra{t-2} U\bra{t-3} \frac{1}{Nc} \sum_{j\in T_{i}} \frac{1}{Nc} \sum_{k\in T_{j}} \frac{1}{Nc} \sum_{l\in T_{k}} \frac{1}{Nc} \sum_{n\in T_l} \sum_{\mu \neq \nu} \xi _{i}^{\mu+1} \xi _{n}^{\mu -3} s_{n}\left( t-3\right).
\label{eq08}
\end{eqnarray}
\end{widetext}
Here, it is easy to show that the third term of Eq. (\ref{eq07}) converge to $\mathcal{N}(0,L_{2}U^{2}\left( t-1\right)U^2\bra{t-2} \alpha / c )$.

Repeating the above expansion up to time $0$, we collect all the correlations caused by the loops and derive the final expression of crosstalk noise at time $t$,
\begin{equation}
Z_i\bra{t} = \mathcal{N}\bra{0,\alpha /c} + \sum_{t_i=0}^{t-1} \mathcal{N}\bra{0,\bra{\prod_{t_i'=t_i}^{t-1}U^2(t_i')} L_{t-t_i} \frac{\alpha}{c}}.
\label{eq09}
\end{equation}
Because each sum in the above equation is independent, one can calculate the variances directly (for proof of independency of each sum, see \cite{Kawamura02,Kitano98}),
\begin{equation}
\sigma^2\bra{t} = \frac{\alpha}{c} + \sum_{t_i=0}^{t-1} \bra{\prod_{t_i'=t_i}^{t-1}U^2(t_i')}L_{t-t_i} \frac{\alpha}{c}.
\label{eq10}
\end{equation}

And we write down explicitly the following closed equations of the order parameter $m(t)$,
\begin{equation}\label{eq:m}
    m\bra{t+1} = \int Dz\left\langle \xi g\bra{\xi m \bra{t} + \sigma \bra{t} z}\right\rangle_\xi,
\end{equation}
\begin{equation}\label{eq:U}
    U\bra{t+1} = \frac{1}{\sigma\bra{t}} \int Dzz \left\langle g \bra{\xi m\bra{t} + \sigma\bra{t} z} \right\rangle_\xi,
\end{equation}
where $\langle\cdot\rangle_\xi$ stands for the average over the retrieval patterns $\xi$, and $Dz = \frac{1}{\sqrt{2\pi}} \exp\bra{-z^2/2}dz$.

We are now ready to present the signal-to-noise treatment of the sequence processing neural networks with narrow-degree-distributed topology at zero temperature. The above equations form a recursive scheme to calculate the dynamical properties of the systems with an arbitrary time step. However, the following practical trouble is how to calculate the loopiness coefficients up to arbitrary order for a given topology.

In the special case of fully-connected networks, $c=1$, all the loopiness coefficient are $L_n = 1$. And substituting into Eq. (\ref{eq10}), we find $\sigma^2$ under the form,
\begin{equation}
\sigma^2\bra{t} = \alpha + U^2(t-1) \sigma^2\bra{t-1}.
\label{eq:fully:sigma}
\end{equation}
This equation, together with Eqs. (\ref{eq:m}-\ref{eq:U}), is coincident with the dynamical equations in \cite{Kawamura02,Amari88}. The stationary equations and simulation results can be found in \cite{Kawamura02}.

\begin{figure}
\includegraphics[width=0.8\textwidth]{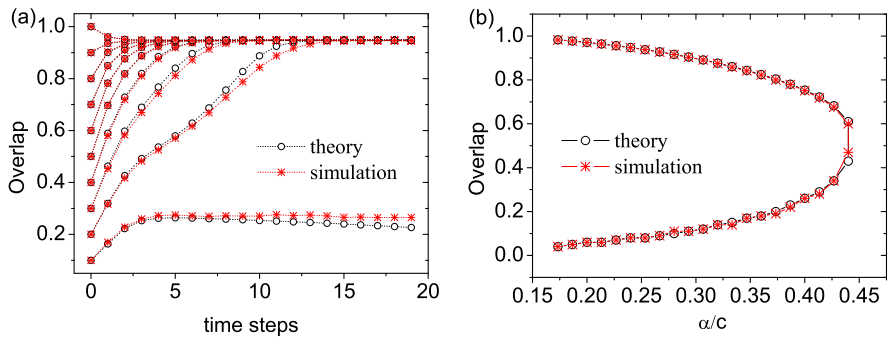}
\caption{(Color online) (a) Temporal evolution of the overlap
parameters of network with random topology, the initial overlap
ranges from $1.0$ to $0.1$ (top to bottom). (b) basin of attraction
of network with random topology. the lower part means the lowest
initial overlap to retrieve successfully. The upper part is the final overlap for successful retrieval. Temperature is zero. The parameters are (a) $N=5000$, $c=0.16$, and $\alpha N=190$; (b) $N=3000$ and $c=0.1$.}
\label{fig2}
\end{figure}

\begin{figure}
\includegraphics[width=0.8\textwidth]{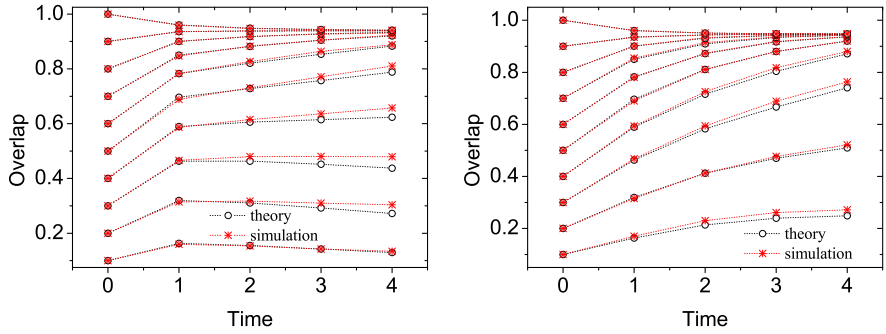}
\caption{(Color online) Temporal evolution of the overlap parameters with (a) regular topology and (b) SW topology. the initial overlap ranges from $1.0$ to $0.1$ (up to down). Temperature is zero. The parameters are $N=5000$, $\bar{k}=800$, and $\alpha N=190$. Rewiring probability of SW network is $0.5$. In our paper, both regular topology and SW topology is in one-dimensional.}
\label{fig3}
\end{figure}

\begin{figure}
\includegraphics[width=0.8\textwidth]{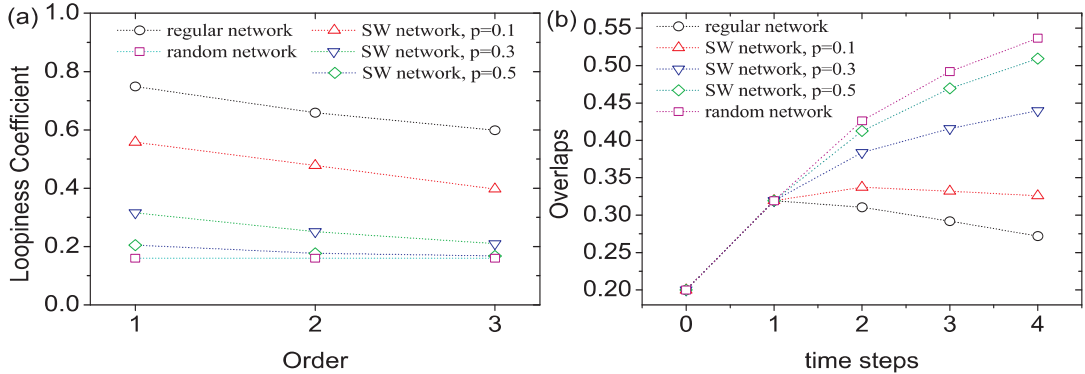}
\caption{(Color online) (a) The first-, second-, and third- order
loopiness coefficients and (b) the theoretical results of $4$ time step dynamics for regular networks, SW networks($p$ is rewiring probability), and random networks. The initial overlaps are $0.2$. The parameters of networks are $N=5000$, $\bar{k}=800$, and $\alpha N=190$. }
\label{fig4}
\end{figure}

For random networks, each node has a probability $c$ to connect with others. So the loopiness coefficients are also $L_n = c$. The temporal evolutions of overlap up to $20$ time steps obtained by both theory and numerical simulations are plotted in Fig. \ref{fig2}. The initial overlap ranges from $0.1$ to $1.0$. Fig. \ref{fig2} shows the basin of attraction from theory and simulations.

However, for regular networks or SW networks, it is impossible to get the exact analytical expressions of the loopiness coefficients. so the only solution is programming. Due to algorithm complexity in large size networks, we only present the first, second, and third order loopiness coefficients. As a result, one only derives the $4$ time step dynamics from Eqs. (\ref{eq10}-\ref{eq:U}). In Fig. \ref{fig3}, we show the theoretical results and simulations of $4$ time step dynamics for regular topology and SW topology with rewiring probability $p=0.5$.

In practice, it was found that the loopiness coefficients decrease with increasing the rewiring probability of SW networks, and the overlaps are increaseed with decreasing the loopiness coefficients (see Fig. \ref{fig4}). From Eq. (\ref{eq10}), the larger loopiness coefficients induce larger variances of the crosstalk noise, and eventually result in a negative effect on performances of neural networks.

Comparing to the sequence processing model, besides of a noise part and a Gaussian part, the local field of Hopfield model has a discrete self-interaction part, and crosstalk noise is no longer normally distributed, which produce studied intractable. But similarly to sequence processing model, the large loopiness coefficient also increases correlations of spin states at different time and leads to large variance of the Gaussian part of crosstalk noise. So it is easy to qualitatively understand why small amount of randomly rewired connections greatly improve the performance of Hopfield model with regular architecture \cite{Patrick03,Torres03,sw1,Morelli04}, and why Hopfield model with random networks is better than SW networks \cite{Patrick03}.

Note that the value of the first-order loopiness coefficient in undirected graph is the number of triples divided by the number of connected triples (See Fig. \ref {fig1}), which is the same with well-known network parameter \textit{clustering coefficient} \cite{Watts98}. Considering Eq. (\ref{eq10}), it shows that $L_1$ plays a much more important role than the higher order loopiness coefficients, and the lower $L_1$ or clustering coefficient make better performance for neural networks. This is indeed confirmed by Kim with Monte-Carlo method to tune the clustering coefficient while retaining the same degree distributions \cite{Kim04}.

We also have to point out that all topologies studied in both theory and simulations are densely-connected networks, our results can not be extended to finite-connectivity networks directly, the reason is: in Eq.(7), using central limit theorem, we assume that the term $\sum_{\mu \neq \nu} \frac{1}{Nc} \sum_{j\in T_{i}} \xi _{i}^{\mu +1} \xi _{j}^{\mu } \hat{s}_{j}^{\mu }\left( t\right)$ converges to a zero mean Gaussian form $\mathcal{N} (0,\alpha / c )$. This assumption is correct in densely-connected networks where $\bar{k}\to\infty$, but incorrect in finite-connectivity networks. However, we find that if $\bar k$ is big($\bar k>=50$), above assumption is a good approximation and our theory gives a good approximation to finite-connectivity network dynamics.

At last, it should be very careful that this theory is not suitable for scale-free networks since the power-law degree distribution is out of action of signal-to-noise analysis and its loopiness coefficients are degree-correlated, which cannot be expressed simply as $L_n$. Attempts of studying dynamics of sequence processing neural networks with power-law and even arbitrary degree distributions have been done by neglecting the loopiness effect \cite{chen07}. Some numerical studies of properties of neural networks with scale-free topology can be found in \cite{stauffer03}.

\section*{Acknowledgments}

This work was supported by the National Natural Science Foundation of China under Grant No. $10305005$ and by the Special Fund for Doctor Programs at Lanzhou University.

\end{document}